\documentclass[preprintnumbers, showpacs, floatfix,onecolumn,preprintnumbers, letterpaper, superscriptaddress,nofootinbib]{revtex4}
\usepackage{amsfonts}
\usepackage{amsmath}
\usepackage{amssymb,epsf}
\usepackage{latexsym}
\usepackage{graphicx,epsfig}
\usepackage{amssymb}
\usepackage{float}
\usepackage{subfigure}
\usepackage{epstopdf}
\usepackage[colorlinks=true,citecolor=blue,linkcolor=blue,urlcolor=black]{hyperref}
\usepackage{dcolumn}
\usepackage{psfrag}
\usepackage{wrapfig}
\usepackage{makeidx}
\usepackage{epsf}

\begin{document}

\title{Conductivity of the one-dimensional holographic $p$-wave
 superconductors in the presence of nonlinear electrodynamics}
\author{Mahya Mohammadi}
\affiliation{Physics Department and Biruni Observatory, Shiraz
University, Shiraz 71454, Iran}
\author{Ahmad Sheykhi}
\email{asheykhi@shirazu.ac.ir} \affiliation{Physics Department and
Biruni Observatory, Shiraz University, Shiraz 71454, Iran}

\affiliation{Research Institute for Astronomy and Astrophysics of
Maragha (RIAAM), University of Maragheh, P. O. Box: 55136-553,
Maragheh, Iran}

\affiliation{Institut f\"{u}r Physik, Universit\"{a}t Oldenburg,
Postfach 2503 D-26111 Oldenburg, Germany}

\begin{abstract}

We investigate analytically as well as numerically the effects of
nonlinear Born-Infeld (BI) electrodynamics on the properties of
$(1+1)$-dimensional holographic $p$-wave superconductor in the
context of gauge/gravity duality. We consider the case in which
the gauge and vector fields backreact on the background geometry.
We apply the Sturm-Liouville eigenvalue problem for the analytical
approach as well as the shooting method for the numerical
calculations. In both methods, we find out the relation between
critical temperature $T_{c}$ and chemical potential $\mu$ and show
that both approaches are in good agreement with each other. We
find that if one strengthen the effect of backreaction as well as
nonlinearity, the critical temperature decreases which means that
the condensation is harder to form. We also explore the
conductivity of the one-dimensional holographic $p$-wave
superconductor for different values of $b$ and $T/T_{c}$. We find
out that the real and imaginary parts of the conductivity have
different behaviors in higher dimensions. The effects of different
values of temperature is more apparent for larger values of
nonlinearity parameter. In addition, for the fixed value of
$T/T_{c}$ by increasing the effect of nonlinearity we observe
larger values for Drude-like peak in real part of conductivity and
deeper minimum for imaginary part.

\end{abstract}

\maketitle

\section{Introduction}
The idea of holographic superconductor was proposed by Hartnoll,
et.al., \cite{H08} through building a holographic $s$-wave
superconductor, in the background of $4$-dimensional Schwarzschild
anti-de Sitter (AdS) black holes. The motivation was to shed light
on the problem of high temperature superconductors. The most
well-known theory of superconductors was proposed by Bardeen,
Cooper and Schrieffer (BCS), which can successfully explain the
mechanism of the low temperature superconductors. According to BCS
theory, the condensation of pairs of electrons  with antiparallel
spins (Cooper pairs) interacting through the exchange of phonon,
into a boson-like state \citep{BCS57}. For building the
holographic superconductor, the correspondence between AdS spaces
and Conformal Field Theory (CFT) plays a crucial role
\cite{H08,H09}. According to the AdS/CFT dictionary, a strong
coupling $d$-dimensional conformal field theory living on the
boundary is equivalent to the weak coupling gravity theory in $(d
+ 1)$-dimensional AdS bulk and each quantity in the bulk has a
dual on the boundary \cite{Maldacena,G98,W98,HR08,R10}. In order
to describe a superconductor at boundary in holographic scenario,
we need a hairy black hole in the bulk. More precisely, we need a
hairy black hole (superconducting phase) for temperatures bellow
the critical value and a black hole with no hair (normal
phase/conductor phase) for upper values. During this process, the
system undergoes the spontaneous $U(1)$ symmetry breaking. The
condensation of a charged operator at the boundary corresponds to
emerge of the hair for black hole in the bulk. The quantum
description of the hair for black hole is the gas of charged
particles which have the same sign charge as the black hole. These
are repelled away by black hole and forbidden to escape to
infinity due to the presence of the negative cosmological constant
in the AdS bulk \cite{H11}. The holographic superconductor theory
grabs a lot of attentions in the past decade (see e.g.
\cite{Hg09,Gu09,HHH08,JCH10,SSh16,SH16,cai15,SHsh(17),
Ge10,Ge12,Kuang13,Pan11,CAI11, SHSH(16),shSh(16),Doa, Afsoon,
cai10,yao13,n4,n5,n6,Gan1}). Moreover, holographic superconductors
have also been explored widely in the regime of nonlinear
electrodynamics (see e.g.
\cite{n4,SH16,SSh16,SHsh(17),SHSH(16),shSh(16),n5,n6,bina,mahya}).
There are several types of nonlinear electrodynamics such as BI
\cite{25}, Exponential \cite{hendi}, Logarithmic \cite{log} and
Power-Maxwell \cite{SH16}, among them the most famous one is the
BI nonlinear electrodynamics which was first proposed for solving
the divergency in the electrical field of the point particles
\cite{25,26,27,28,29}. The studies on the holographic
superconductors have also generalized to other types such as
$p$-wave superconductors. The $p$-wave superconductivity is a
phase of matter where produces when the electrons are bounded with
parallel spins by exchange of the electronic excitations with
angular momentum $\ell = 1$ and condense in a triplet state. The
terms odd parity superconductivity, $p$-wave superconductivity and
triplet superconductivity all are equivalent \cite{superp}.
Various models of holographic $p$-wave superconductors have been
investigated. Holographic $p$-wave superconductors can be studied
by condensation of a charge vector field in the bulk which
corresponds to the vector order parameter in the boundary \cite
{Caip,cai13p}. This implies that the spin-$1$ order parameter can
be corresponded to the condensation of a $2$-form field in the
gravity side\cite{Donos}. In \cite{Gubser}, this type of
holographic superconductors characterized by introducing a $SU(2)$
Yang-Mills gauge field in the bulk which one of the gauge degrees
of freedom corresponds to the vector order parameter at the
boundary. In addition, an alternative method to describe this kind
of holographic superconductor emerges by adopting a complex vector
field charged under a $U(1)$ gauge field which is equivalent to a
strongly coupled system involving a charged vector operator with a
global $U(1)$ symmetry at the boundary \cite{chaturverdip15}. For
this type of holographic superconductor, by decreasing temperature
below the critical value, the normal phase becomes unstable and we
observe the formation of vector hair which corresponds to
superconducting phase. Other investigations on the holographic
$p$-wave superconductors have been carried out in
(e.g.\cite{Roberts8,zeng11,cai11p,pando12,momeni12p,gangopadhyay12,chaturverdip15}).

Furthermore, the $(1+1)$-dimensional holographic superconductors
have been developed in the background of BTZ black hole
\cite{R10}. BTZ black hole is the well-known solutions of general
relativity in $(2+1)$-dimensional spacetime which plays a crucial
role in understanding the gravitational interaction in low
dimensional spacetimes. In order to study the one-dimensional
holographic superconductor one may apply $AdS_3/CFT_2$
correspondence \cite{Car1,Ash,Sar,Wit1,Car2}. One-dimensional
holographic $s$-wave and $p$-wave superconductors were analyzed
both analytically and numerically from different points of view in
(see e.g.
\cite{Bu,Wang,chaturvedi,L12,momeni,peng17,lashkari,hua,yanyan,yan,bina,mahya,mahyap}).
It is worth noting that most investigations on the
$(1+1)$-dimensional holographic $p$-wave superconductors are done
in the framework of linear Maxwell electrodynamics. Therefore, it
is fascinating to study the effects of nonlinearity in such a
holographic superconductor. In the present work, we would like to
extend the investigation on the one-dimensional holographic
superconductor by considering the BI nonlinear electrodynamics
when gauge and vector fields backreact on background geometry. It
is worth noting that the $(1+1)$-dimensional holographic $p$-wave
superconductor is located on the boundary of $(2+1)$ dimensional
spacetime. While the gauge and the vector fields are defined in
the $(2+1)$-dimensional BTZ black hole. It is well-known that the
electric field of a point charge in $(2+1)$-dimension has the form
$E(r)=q/r$. Thus, there is still a divergency in the electric
field at the location of point charge ($r=0$). However, taking the
BI electrodynamics into account can remove this divergency as
well. This is the main motivation for investigating the
$(1+1)$-dimensional holographic $p$-wave superconductor in the
presence of nonlinear BI electrodynamics.

We shall employ both analytical and numerical approaches. Our
analytical study is based on the Sturm-Liouville eigenvalue
problem while the shooting method is used for the numerical
calculations. Our aim is to find a relation between the critical
temperature $T_{c}$ and the chemical potential $\mu$ for different
values of backreaction and nonlinear parameters. We also
investigate the effects of nonlinearity on the real and imaginary
parts of conductivity.

This article is organized as follow. In section \ref{sec2} we
introduce the one-dimensional holographic $p$-wave superconductor.
In section \ref{sec3}, we study condensation of the vector field
both analytically and numerically. In section \ref{sec4}, we
calculate the critical exponent of this type of holographic
superconductor analytically as well as numerically. In section
\ref{sec5}, we explore the holographic conductivity for this
model. Finally, in section \ref{sec6} we present a summary of our
results and discussion.

\section{The holographic p-wave model}\label{sec2}
In a three dimensional spacetime, the action of Einstein gravity
in the presence of nonlinear BI electrodynamics and negative
cosmological constant can be written as
\begin{eqnarray}
&&S =\frac{1}{2\kappa ^{2}}\int d^{3}x\sqrt{-g} \mathcal{L}_{G}%
 +\int d^{3}x\sqrt{-g}\mathcal{L}_{m}, \notag \\
&& \mathcal{L}_{G}= R+\frac{2}{l^{2}} , \    \   \  \
\mathcal{L}_{m}=
\mathcal{L}(\mathcal{F})-\frac{1}{2}\rho_{\mu\nu}^{\dagger}
\rho^{\mu\nu}-m^{2} \rho_{\mu}^{\dagger} \rho^{\mu} + i q \gamma
\rho_{\mu} \rho_{\nu}^{\dagger} F^{\mu\nu} .\label{act}
\end{eqnarray}%
In the Lagrangian of the matter field, $\mathcal{L}_{m}$, the
constants $m$ and $q$ are the mass and charge of the vector field
$\rho_{\mu}$, respectively. Here, $\kappa ^{2}=8\pi G_{3}$ where
$G_{3}$ is the $3$-dimensional Newtonian gravitation constant in
the bulk. In addition, the metric determinant, Ricci scalar and
AdS radius are characterized by $g$, $R$ and $l$, respectively.
Also, $F_{\mu\nu}=\nabla_{\mu} A_{\nu}-\nabla_{\nu} A_{\mu}$ is
the strength of the Maxwell field with $A_{\mu}$ as the vector
potential. Considering $D_{\mu}=\nabla_{\mu}- i q A_{\mu}$ we can
define $\rho_{\mu\nu}=D_{\mu} \rho_{\nu}-D_{\nu} \rho_{\mu}$. The
last term in the matter Lagrangian, which represents a nonlinear
interaction between $\rho_{\mu}$ and $A_{\mu}$ with $\gamma$ as
the magnetic moment, can be ignored in the present work because we
consider the case without external magnetic field. The Lagrangian
density of the BI nonlinear
electrodynamics is given by $%
\mathcal{L}(\mathcal{F})$ can be defined as %
\begin{equation}
\mathcal{L}(\mathcal{F})=\frac{1}{b}\left( 1-\sqrt{1+\frac{b\mathcal{F}}{2}}%
\right) ,
\end{equation}%
where $b$ is the nonlinear parameter and $\mathcal{F}=F_{\mu\nu}
F^{\mu\nu}$. When $b\rightarrow 0$, $\mathcal{L}(\mathcal{F})$
reduces to  the standard Maxwell Lagrangian.

Varying the action (\ref{act}) with respect to the metric
$g_{\mu\nu}$, the gauge field $A_{\mu}$ and the vector field
$\rho_{\mu}$, yields the equations of motion for the gravitational
and the bulk matter fields as
\begin{eqnarray}
\frac{1}{2\kappa ^{2}}\left[ R_{\mu \nu }-g_{\mu \nu }\left( \frac{R}{2}%
+\frac{1}{l^{2}}\right) \right]   &=& %
- 2 F_{\mu \lambda}F_{\nu}{}^{\lambda}\mathcal{L}_{\mathcal{F}}+\frac{1}{2}\mathcal{L}_{m} g_{\mu \nu } %
+\frac{1}{2}\left[ \rho ^{\dagger }{}_{\mu \lambda} \rho _{\nu }^{\lambda }+m^2 \rho ^{\dagger }{}_{\mu }
  \rho _{\nu } -i \gamma  q F_{\nu }^{\lambda } \left(\rho _{\mu } \rho ^{\dagger }{}_{\lambda }-
  \rho ^{\dagger }{}_{\mu } \rho _{\lambda }\right)+\mu \leftrightarrow \nu
  \right], \nonumber\\
\label{Eein}
\end{eqnarray}%
\begin{equation}\label{eqmax}
\nabla ^{\nu }\left(- 4\mathcal{L}_{\mathcal{F}}F_{\nu \mu }\right) =i q \left(\rho ^{\nu }
 \rho ^{\dagger }{}_{\nu  \mu }- \rho ^{\nu  \dagger } \rho _{\nu  \mu }\right)+
 i q \gamma \nabla^{\nu} \left(\rho _{\nu } \rho ^{\dagger }{}_{\mu }-\rho ^{\dagger }{}_{\nu } \rho _{\mu } \right),
\end{equation}
\begin{equation}\label{eqvector}
D ^{\nu } \rho_{\nu \mu }-m^{2} \rho_{\mu }+ i q \gamma \rho^{\nu
} F_{\nu\mu} =0,
 \end{equation}
where, $\mathcal{L}_{\mathcal{F}}=\partial \mathcal{L}/\partial
{\mathcal{F}}$. In order to study backreacted $(1+1)$-dimensional
holographic $p$-wave superconductor in the presence of BI
nonlinear electrodynamics, we consider the following ansatz for
the metric and bulk fields
\begin{eqnarray} \label{metric}
&&{ds}^{2}=-f(r)e^{-\chi (r)}{dt}^{2}+\frac{{dr}^{2}}{f(r)}+r^{2}{dx}^{2}%
,\\
&& \rho_{\nu} dx^{\nu}=\rho_x(r) dx, \  \   \  \   A_{\nu}
dx^{\nu}=\phi (r) dt,  \label{rhoA}
\end{eqnarray}%
where the $x$-component of the vector field $\rho_{x}$ corresponds
to the expectation value $\langle J_{x}\rangle$ which plays the
role of the order parameter in the boundary theory. The Hawking
temperature of the black hole is given by \cite{mahya}
\begin{equation}
T=\frac{e^{-\chi (r_{+})/2}f^{^{\prime }}(r_{+})}{4\pi }. \label{temp}
\end{equation}%
Substituting relations (\ref{metric}) and (\ref{rhoA}) in the
field equations (\ref{Eein}) and (\ref{eqmax}), we arrive at
\begin{equation}\label{eqfr}
f'(r)-\frac{2 r}{l^2}+2 \kappa ^2 r \left[\frac{1}{b \sqrt{1-b
e^{\chi (r)} \phi '^2(r)}}-\frac{1}{b}+\frac{q^2 \rho_{x}^2(r)
e^{\chi (r)} \phi^2 (r)}{r^2 f(r)}+ \frac{f(r)
\rho_{x}'^2(r)}{r^2}+\frac{m^2 \rho_{x}^2(r)}{r^2}\right]=0,
\end{equation}
\begin{equation}\label{eqchir}
\chi '(r)+\frac{4 \kappa ^2}{r} \left[\frac{q^2 \rho_{x}^2 (r) e^{\chi (r)} \phi ^2 (r)}{f^2 (r)}+\rho_{x}'^2 (r)\right]=0,
\end{equation}
\begin{equation}\label{eqphir}
\phi ''(r)+\phi '(r) \left[\frac{\chi '(r)}{2}+\frac{1}{r}-\frac{b e^{\chi (r)} \phi '^2(r)}{r}\right]
-\frac{2 q^2 \rho_{x}^2(r) \phi (r)}{r^2 f(r)} \left[1-b e^{\chi (r)} \phi '^2(r)\right]^{3/2}=0,
\end{equation}
\begin{equation}\label{eqrhor}
\rho_{x}''(r)+\rho_{x}'(r) \left[\frac{f'(r)}{f(r)}-\frac{\chi '(r)}{2}-\frac{1}{r}\right]
+\rho_{x}(r) \left[\frac{q^2 e^{\chi (r)} \phi ^2(r)}{f^2(r)}-\frac{m^2}{f(r)}\right]=0.
\end{equation}

Here, the prime denotes derivative with respect to $r$. In the
presence of the nonlinear BI electrodynamics the Eqs.
(\ref{eqrhor}) and (\ref{eqchir}) do not change in comparison with
the linear Maxwell case. In the limiting case where $b\rightarrow
0$ the equations of motion of the Maxwell field are reproduced
\cite{mahyap}. If we consider the probe limit by setting
$\kappa=0$, the equations of motion (\ref{eqphir}) and
(\ref{eqrhor}) turn to the corresponding equations in
\cite{alkac}. There are scaling symmetries of the equations of motion (\ref{eqfr})-(\ref{eqrhor}) that we can use to
set $q$ and $l$ equal to unity.
\begin{gather}
q\rightarrow q/a,\text{ \ \ \ \ }\phi \rightarrow a\phi ,\text{ \ \ \ \ }%
\rho_{x} \rightarrow a\rho_{x} , \   \   \   \kappa \rightarrow \kappa /a, \   \   \   b\rightarrow b /a^{2},\\
\notag \\
l\rightarrow al,\text{ \ \ \ \ }r\rightarrow ar,\text{ \ \ \ \ }q\rightarrow
q/a, \text{ \ \ \ \ }b\rightarrow b a^{2}, \   \   \
m\rightarrow m/a.
\end{gather}%
In addition, there is another symmetry which leaves the metric unchanged
\begin{equation}%
e^{\chi}\rightarrow a^{2}e^{\chi},\text{ \ \ \ \ } t\rightarrow at,\text{ \ \ \ \ } \phi \rightarrow a^{-1}\phi \label{chisym},
\end{equation}
based on Eq. (\ref{chisym}) the boundary value of
$\chi$ is constant, so we can set $\chi =0$. Considering
$\mu$ and $\rho$ as the chemical potential and charge density, the
asymptotic behavior $(r\rightarrow\infty)$ of the field equations
are given by
\begin{equation}\label{eqasymp}
\phi(r) \sim \rho+\mu \ln(r), \  \   \  f(r)\sim r^{2},\  \
\chi(r)\rightarrow0, \  \   \  \rho _x(r) \sim \frac{\rho
_{x_-}}{r^{-m}}+\frac{\rho _{x_+}}{r^{+m}}.
\end{equation}%
Taking the Breitenlohner-Freedman (BF) bound into account,
$m^{2}\geq 0$, $\rho _{x_-}$ plays the role of source and $\rho
_{x_+}$ known as $x$-component of the expectation value of the
order parameter $\langle J_{x} \rangle$\cite{wen18}. Hereafter we
set $m^{2}=1$.  In the next sections, we investigate, analytically
as well as numerically, the properties of one-dimensional
backreacted holographic $p$-wave superconductor in the presence of
BI nonliner electrodynamics.

\begin{figure*}[t]
\centering
\subfigure[~$b=0$]{\includegraphics[width=0.3\textwidth]{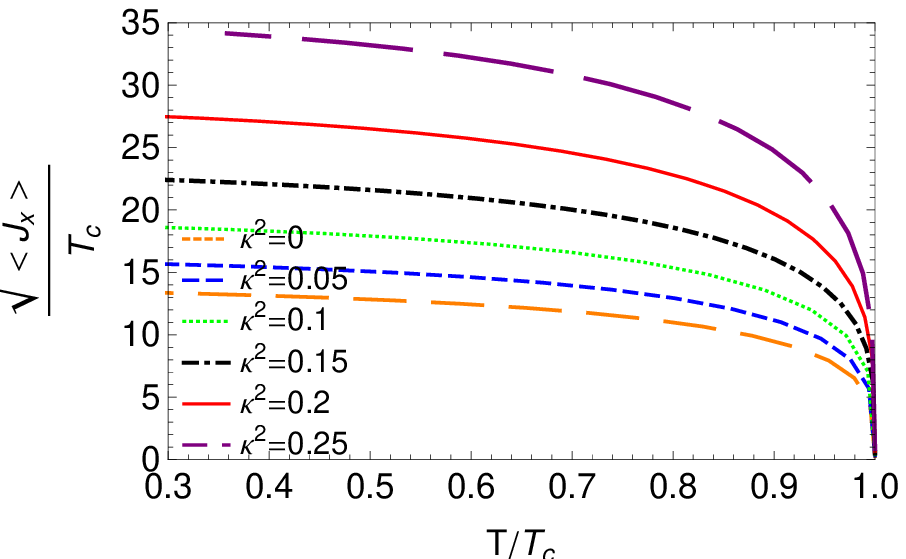}} \qquad %
\subfigure[~$b=0.04$]{\includegraphics[width=0.3\textwidth]{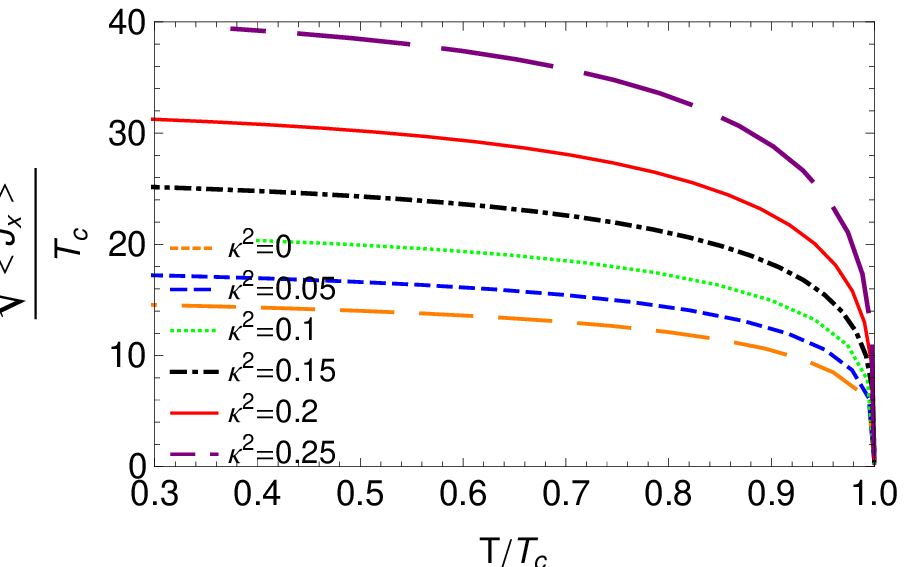}} \qquad %
\subfigure[~$b=0.08$]{\includegraphics[width=0.3\textwidth]{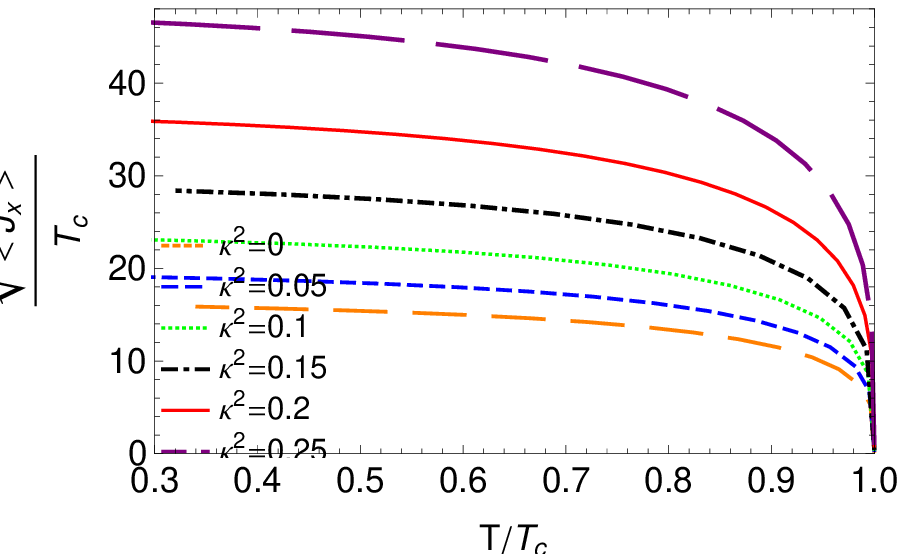}}
\caption{The behavior of condensation parameter as a function of temperature for different values of backreaction.}
\label{fig1}
\end{figure*}

\begin{figure*}[t]
\centering
\subfigure[~$  \kappa^{2}=0$]{\includegraphics[width=0.3\textwidth]{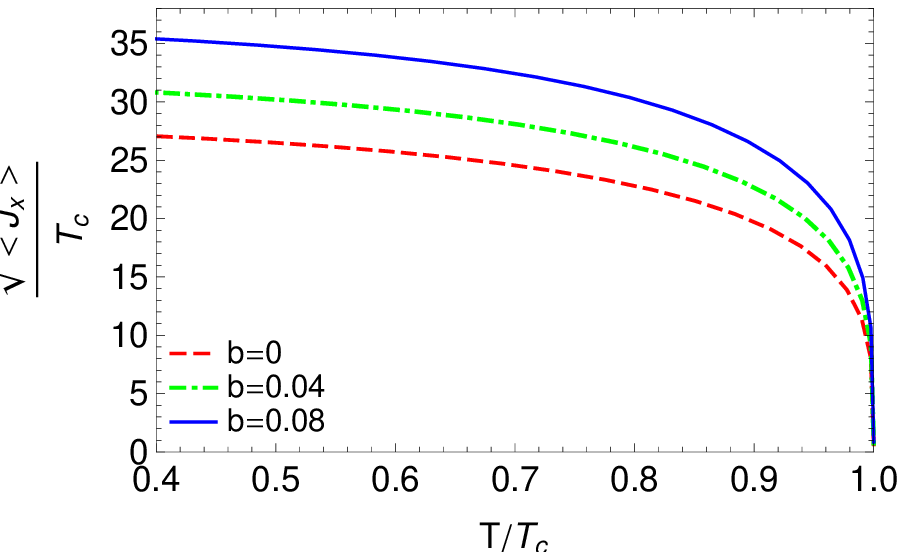}} \qquad %
\subfigure[~$  \kappa^{2}=0.10$]{\includegraphics[width=0.3\textwidth]{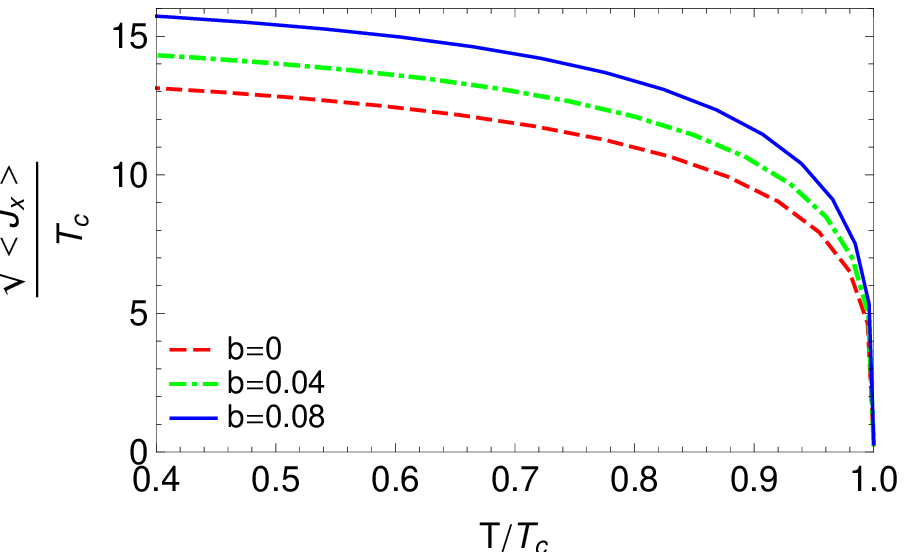}} \qquad %
\subfigure[~$  \kappa^{2}=0.20$]{\includegraphics[width=0.3\textwidth]{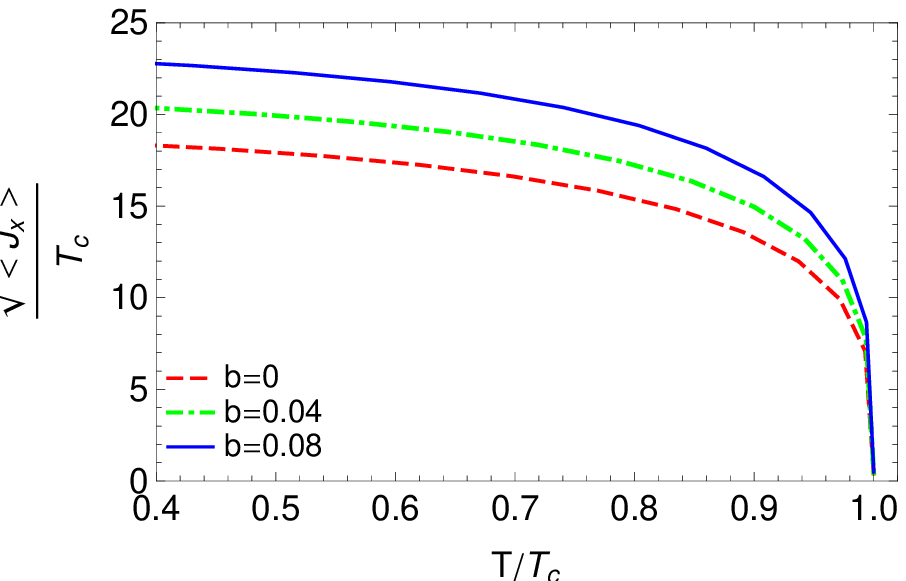}}
\caption{Plot of condensation as a function of temperature with $ m^{2}=1 $ for different values of the nonlinearity parameter $ b $ .}
\label{fig3}
\end{figure*}
\begin{table*}[t]
\label{tab1}
\begin{center}
\begin{tabular}{c|c|c|c|c|c|c|}
\cline{2-3}\cline{2-7}\cline{4-7}
& \multicolumn{2}{|c|}{$b=0$} & \multicolumn{2}{|c|}{$b=0.04$} &
\multicolumn{2}{|c|}{$b=0.08$} \\ \cline{2-3}\cline{2-7}\cline{4-7}
& Analytical & Numerical & Analytical & Numerical & Analytical & Numerical
\\ \hline
\multicolumn{1}{|c|}{$\kappa ^{2}=0$} & $0.0478$ & $0.0503 $ & $0.0416$ & $%
0.0454$ & $0.0343$ & $0.0406$ \\ \hline
\multicolumn{1}{|c|}{$\kappa ^{2}=0.05$} & $0.0443$ & $0.0410$ & $0.0379$ & $%
0.0366$ & $0.0303$ & $0.0324$ \\ \hline
\multicolumn{1}{|c|}{$\kappa ^{2}=0.1$} & $0.0424$ & $0.0330$ & $0.0361$ & $%
0.0290$ & $0.0281$ & $0.0254$ \\ \hline
\multicolumn{1}{|c|}{$\kappa ^{2}=0.15$} & $0.0394$ & $ 0.0260 $ & $0.0331$ & $%
0.0226$ & $0.0248$ & $0.0195$ \\ \hline
\multicolumn{1}{|c|}{$\kappa ^{2}=0.2$} & $0.0353$ & $0.0201 $ & $0.0291$ & $%
0.0172$ & $0.0205$ & $0.0146$ \\ \hline
\multicolumn{1}{|c|}{$\kappa ^{2}=0.25$} & $0.0302$ & $0.0152$ & $0.0241$ & $%
0.0127$ & $0.0153$ & $0.0106$ \\ \hline
\end{tabular}%
\caption{Analytical and Numerical results of $ T_{c}/ \mu $ for
different values of the backreaction and nonlinear parameters with
trial function $F(z)=1-\alpha z^2$.}
\end{center}
\end{table*}
\begin{table*}[t]
\label{tab2}
\begin{center}
\begin{tabular}{c|c|c|c|c|c|c|}
\cline{2-3}\cline{2-7}\cline{4-7}
& \multicolumn{2}{|c|}{$b=0$} & \multicolumn{2}{|c|}{$b=0.04$} &
\multicolumn{2}{|c|}{$b=0.08$} \\ \cline{2-3}\cline{2-7}\cline{4-7}
& Analytical & Numerical & Analytical & Numerical & Analytical & Numerical
\\ \hline
\multicolumn{1}{|c|}{$\kappa ^{2}=0$} & $0.0466$ & $0.0503 $ & $0.0400$ & $%
0.0454$ & $0.0320$ & $0.0406$ \\ \hline
\multicolumn{1}{|c|}{$\kappa ^{2}=0.05$} & $0.0397$ & $0.0410$ & $0.0347$ & $%
0.0366$ & $0.0243$ & $0.0324$ \\ \hline
\multicolumn{1}{|c|}{$\kappa ^{2}=0.1$} & $0.0375$ & $0.0330$ & $0.0328$ & $%
0.0290$ & $0.0215$ & $0.0254$ \\ \hline
\multicolumn{1}{|c|}{$\kappa ^{2}=0.15$} & $0.0342$ & $ 0.0260 $ & $0.0298$ & $%
0.0226$ & $0.0178$ & $0.0195$ \\ \hline
\multicolumn{1}{|c|}{$\kappa ^{2}=0.2$} & $0.0296$ & $0.0201 $ & $0.0258$ & $%
0.0172$ & $0.0132$ & $0.0146$ \\ \hline
\multicolumn{1}{|c|}{$\kappa ^{2}=0.25$} & $0.0238$ & $0.0152$ & $0.0208$ & $%
0.0127$ & $0.0083$ & $0.0106$ \\ \hline
\end{tabular}%
\caption{Analytical and Numerical results of $ T_{c}/ \mu $ for
different values of the backreaction and nonlinear parameters with $F(z)=1-\alpha z^3$.}
\end{center}
\end{table*}

\begin{table*}[t]
\label{tab3}
\begin{center}
\begin{tabular}{c|c|c|c|c|c|c|}
\cline{2-3}\cline{2-7}\cline{4-7}
& \multicolumn{2}{|c|}{$b=0$} & \multicolumn{2}{|c|}{$b=0.04$} &
\multicolumn{2}{|c|}{$b=0.08$} \\ \cline{2-3}\cline{2-7}\cline{4-7}
& Analytical & Numerical & Analytical & Numerical & Analytical & Numerical
\\ \hline
\multicolumn{1}{|c|}{$\kappa ^{2}=0$} & $0.0457$ & $0.0503 $ & $0.0387$ & $%
0.0454$ & $0.0303$ & $0.0406$ \\ \hline
\multicolumn{1}{|c|}{$\kappa ^{2}=0.05$} & $0.0379$ & $0.0410$ & $0.0328$ & $%
0.0366$ & $0.0205$ & $0.0324$ \\ \hline
\multicolumn{1}{|c|}{$\kappa ^{2}=0.1$} & $0.0356$ & $0.0330$ & $0.0308$ & $%
0.0290$ & $0.0171$ & $0.0254$ \\ \hline
\multicolumn{1}{|c|}{$\kappa ^{2}=0.15$} & $0.0321$ & $ 0.0260 $ & $0.0278$ & $%
0.0226$ & $0.0131$ & $0.0195$ \\ \hline
\multicolumn{1}{|c|}{$\kappa ^{2}=0.2$} & $0.0274$ & $0.0201 $ & $0.0236$ & $%
0.0172$ & $0.0084$ & $0.0146$ \\ \hline
\multicolumn{1}{|c|}{$\kappa ^{2}=0.25$} & $0.0214$ & $0.0152$ & $0.0186$ & $%
0.0127$ & $0.0039$ & $0.0106$ \\ \hline
\end{tabular}%
\caption{Analytical and Numerical results of $ T_{c}/ \mu $ for
different values of the backreaction and nonlinear parameters with $F(z)=1-\alpha z^4$.}
\end{center}
\end{table*}
\section{condensation of vector field}\label{sec3}
Let us now investigate the relation between the critical
temperature $T_{c}$ and the chemical potential $\mu$ for
holographic $p$-wave superconductor. In particular, we would like
to explore the effects of backreaction as well as nonlinearity
parameter on the critical temperature.
\section*{Analytical approach}
In order to follow our analytical studies, we apply the
Sturm-Liouville eigenvalue problem and define $z=r_{+}/r$ as a new
variable where $0\leq z \leq1$. Therefore, Eqs.
(\ref{eqfr})-(\ref{eqrhor}) turn to
\begin{equation}\label{eqfz}
f'(z)+\frac{2 r_+^2}{z^3}+\frac{2 \kappa ^2}{z} \left[- \rho _x{}^2(z)-\frac{ e^{\chi (z)} \phi ^2 (z) \rho _x{}^2(z)}{f(z)}-\frac{z^4 f(z) \rho _x'{}^2(z)}{r_+^2}+\frac{r_+^2}{b z^2} \left[1-\frac{1}{\sqrt{1-\frac{b z^4 e^{\chi (z)} \phi '^2(z)}{r_+^2}}}\right]\right]=0,
\end{equation}
\begin{equation}\label{eqchiz}
\chi '(z)-4 \kappa ^2 \left[\frac{e^{\chi (z)} \phi ^2(z) \rho _x{}^2(z)}{z f^2(z)}+\frac{z^3 \rho _x'{}^2(z)}{r_+^2}\right]=0,
\end{equation}
\begin{equation}\label{eqphiz}
\phi ''(z)+\phi '(z) \left[\frac{1}{z}+\frac{\chi '(z)}{2}+\frac{b z^3 e^{\chi (z)} \phi '^2(z)}{r_+^2}\right]-\frac{2 \phi (z) \rho _{x} ^2 (z)} {z^2 f(z)}\left[1-\frac{b z^4 e^{\chi (z)} \phi '^2(z)}{r_+^2}\right]^{3/2}=0,
\end{equation}
\begin{equation}\label{eqrhoz}
\rho _x''(z)+\rho _x'(z) \left[\frac{3}{z}-\frac{\chi '(z)}{2}+\frac{f'(z)}{f(z)}\right] +\rho _x(z) \left[\frac{ r_+^2 e^{\chi (z)} \phi ^2(z)}{z^4 f^2(z)}-\frac{ r_+^2}{z^4 f(z)}\right]=0.
\end{equation}
Here, the prime indicates the derivative with respect to $z$. In
the vicinity of the critical temperature the expectation value of
$\langle J_x \rangle$ is tiny so we take it as an expansion
parameter
\begin{equation*}
\epsilon \equiv \left\langle J_{x}\right\rangle.
\end{equation*}%
We concentrate on the solutions for small values of the
condensation parameter $\epsilon $, because near the critical
temperature we have $\epsilon\ll 1$. Therefore, we expand the
functions in terms of the $\epsilon$ as
\begin{gather*}
f\approx f_{0}+\epsilon ^{2}f_{2}+\epsilon ^{4}f_{4}+\cdots , \\
\chi \approx \epsilon ^{2}\chi _{2}+\epsilon ^{4}\chi _{4}+\cdots , \\
\phi \approx \phi _{0}+\epsilon ^{2}\phi _{2}+\epsilon ^{4}\phi _{4}+\cdots , \\
\rho_{x} \approx \epsilon \rho_{x_{1}}+\epsilon
^{3}\rho_{x_{3}}+\epsilon ^{5}\rho _{x_{5}} +\cdots.
\end{gather*}%
Moreover, the chemical potential can be expressed as
\begin{equation*}
\mu =\mu _{0}+\epsilon ^{2}\delta \mu _{2}+...\rightarrow \epsilon
\thickapprox \Bigg(\frac{\mu -\mu _{0}}{\delta \mu
_{2}}\Bigg)^{1/2}, \    \  \   \  \delta \mu _{2}>0.
\end{equation*}%
Near the phase transition $\mu _{c}=\mu _{0}$, so the order
parameter vanishes. In addition, the mean field value for the
critical exponent $\beta =1/2$ is obtained.
\\At zeroth order of $\epsilon$, the equation for the gauge field (\ref{eqphiz}) turns to
\begin{equation}\label{eqphiasym}
\phi ''(z)+\frac{\phi '(z)}{z}+\frac{b z^3 \phi
'^3(z)}{r_{+}^2}=0.
\end{equation}
We can find the solution for this equation as fallow
\begin{equation}\label{eqphiiians}
\phi(z)=\lambda  r_{+} \log (z)-\frac{1}{4} b \lambda ^3 r_{+} \left(z^2-1\right), \  \  \ \lambda =\frac{\mu }{r_+}.
\end{equation}
Substituting solution (\ref{eqphiiians}) into Eq. (\ref{eqfz}), we
arrive at
\begin{equation}\label{phi0}
f'(z)+\frac{2 r_+^2}{z^3}+\frac{2 \kappa ^2 r_+^2}{b
z^3}\left(1-\frac{1}{\sqrt{1-\frac{b z^4 \phi
'^2(z)}{r_+^2}}}\right)=0,
\end{equation}
The solution for $f(z)$, at zeroth order of $\epsilon$, can be
obtained as
\begin{equation}
f(z)=\frac{r_+^2 g(z)}{z^2}, \    \   \  \ g(z)= 1-z^2+\kappa ^2
\lambda ^2 z^2 \log (z)+\frac{1}{8} b \kappa ^2 \lambda ^4 z^2
-\frac{1}{8} b \kappa ^2 \lambda ^4 z^4.  \label{eqfff}
\end{equation}%
Near the boundary, we can define the function $\rho _{x}(z)$ based
on the trial function  $F(z)=1-\alpha z^{\varpi}$ in which $\varpi\geq2$ and satisfies the
boundary conditions $F(0)=1$ and $F^{^{\prime }}(0)=0$,
\begin{equation}
\rho _{x}(z)=\frac{\langle J_{x}\rangle }{\sqrt{2}r_{+}^{\Delta }}z^{\Delta
}F(z).  \label{eqrhoo}
\end{equation}%
Inserting Eqs. (\ref{eqfff}) and (\ref{eqrhoo}) in Eq.
(\ref{eqrhoz}), we arrive at
\begin{equation}
F''(z)+F'(z) \left[\frac{g'(z)}{g(z)}+\frac{2 \Delta
}{z}+\frac{1}{z}\right]+F(z) \left[\frac{\Delta g'(z)}{z
g(z)}-\frac{1}{z^2 g(z)}+\frac{\Delta ^2}{z^2}\right]-\frac{F(z)
\lambda ^2 \log (z) }{2 g^2(z)}\left[b \lambda ^2 r_{+}
\left(z^2-1\right)-2 \log (z)\right]=0.
\end{equation}\label{eqF}
The Sturm-Liouville form of this equation is%
\begin{equation}
\left[ T(z)F^{\prime }(z)\right] ^{\prime }+P(z)T(z)F(z)+\lambda
^{2}Q(z)T(z)F(z)=0,  \label{sl}
\end{equation}%
where
\begin{equation}\label{eqtz}
T(z)= z^{2 \Delta +1} \left[\left(1-z^2\right) \left(1+\frac{b}{8} \kappa ^2 \lambda ^4 z^2\right)+ \kappa ^2 \lambda ^2 z^2 \log (z)\right],
\end{equation}
\begin{equation}\label{eqpz}
P(z)=\frac{\Delta}{z}  \left(\frac{g'(z)}{g(z)}+\frac{\Delta }{z}\right)-\frac{1}{z^2 g(z)},
\end{equation}
\begin{equation}\label{eqqz}
Q(z)=\frac{\log (z)}{ g^2(z)}\left[ \log (z)+\frac{b}{2} \lambda ^2 r_+ \left(1-z^2\right)\right] .
\end{equation}
According to the Sturm-Liouville eigenvalue problem, we should
minimize the following expression with respect to $\alpha $.
\begin{equation}
\lambda ^{2}=\frac{\int_{0}^{1}T\left( F^{\prime 2}-PF^{2}\right) dz}{%
\int_{0}^{1}TQF^{2}dz}, \label{l2}
\end{equation}%
The definition of the backreaction parameter, based on the
iteration method, is \cite{LPJW2015}
\begin{equation}
\kappa _{n}=n\Delta \kappa ,\ \ \ n=0,1,2,\cdots ,\   \   \Delta \kappa =\kappa _{n+1}-\kappa _{n}.
\end{equation}%
where we take $\Delta\kappa =0.05$. In addition, we have%
\begin{gather}
\kappa ^{2}\lambda ^{2}={\kappa _{n}}^{2}\lambda ^{2}={\kappa _{n}}%
^{2}(\lambda ^{2}|_{\kappa _{n-1}})+O[(\Delta \kappa )^{4}],\  \  \kappa _{-1}=0 ,\  \ \lambda ^{2}|_{\kappa _{-1}}=0,\\
b\lambda ^{2}=b\left( \lambda ^{2}|_{b=0}\right) +\mathcal{O}(b^{2}).
\end{gather}%
Using Eqs. (\ref{temp}) and (\ref{eqfr}), the critical
temperature, at zeroth order with respect to $\epsilon$, is given
by
\begin{equation}\label{eqtemp}
T_c=\frac{f'\left(r_{+c}\right)}{4 \pi }=\frac{r_{+c}}{4 \pi }
\left[2-\kappa ^2 \lambda ^2+\frac{1}{4} b \kappa ^2 \lambda
^4\right]=\frac{1}{4\pi }\left(\frac{\mu }{\lambda
}\right)\left[2-\kappa _{n}^{2}(\lambda ^{2}|_{\kappa
_{n-1}})+\frac{1}{4} b\kappa _{n}^{2}(\lambda ^{4}|_{\kappa
_{n-1},b=0})\right].
\end{equation}
Considering three different forms of the trial function
$F(z)$, the analytical results of $T_{c}/ \mu$ affected by
different values of backreaction and nonlinear parameters are
listed in tables I, II and III. Based on these results, the
effects of increasing the backreaction parameter $\kappa$ for a
fixed values of nonlinear parameter $b$ are the same as increasing
the nonlinear parameter for a fixed value of $\kappa$. In other
words, in both cases, the value of $T_{c}/ \mu$ decreases by
increasing the backreaction or nonlinear parameters. Thus, the
presence of backreaction and BI nonlinear electrodynamics makes
the vector hair harder to form. In addition, for the case with
$b=0$, the results of \cite{mahyap} for $T_{c}/ \mu$ are
reproduced.

\section*{Numerical solution}
To do our numerical solution for the $(1+1)$-dimension holographic
$p$-wave superconductor in the presence of backreaction and BI
nonlinear electrodynamics, we employ the shooting method. For this
purpose, we need to know the behavior of the equations of motion
(\ref{eqphiz})-(\ref{eqchiz}) both at horizon and boundary. We use
the facts that $\phi(z=1)=0$, otherwise the norm of the gauge
field $A_{\mu}$ will be ill-defined at the horizon where
$f(z=1)=0$. By using these conditions, we can expand the metric
functions and vector field, around $z=1$, as
\begin{gather}
f(z)=f_{1}\left( 1-z\right) +f_{2}\left( 1-z\right) {}^{2}+\cdots , \\
\chi (z)=\chi _{0}+\chi _{1}\left( 1-z\right) +\chi _{2}\left( 1-z\right)
{}^{2}+\cdots , \\
\phi (z)=\phi _{1}\left( 1-z\right) +\phi _{2}\left( 1-z\right)
{}^{2}+\cdots , \\
\rho_{x} (z)=\rho_{x_{0}}+\rho_{x_{1}} \left( 1-z\right)
+\rho_{x_{2}}\left( 1-z\right) {}^{2}+\cdots.
\end{gather}%
The higher orders will be disregarded because in the vicinity of
horizon $(1-z)^{n}$ is very small and can be neglected. In this
method, we can write all coefficients in terms of $\phi _{1}$,
$\rho _{x_{0}}$ and $\chi _{0}$. By varying these three parameters
at the horizon, we try to gain the desirable state $\rho
_{x_{-}}(\infty)=\chi(\infty) =0$. In addition, we can set
$r_{+}=1$ by virtue
of the equations of motion's symmetry%
\begin{equation*}
r\rightarrow ar,\text{ \ \ \ \ }f\rightarrow a^{2}f,\text{ \ \ \ \ }\phi
\rightarrow a\phi .
\end{equation*}
Consequently, the numerical values of $T_{c}/ \mu$ for different
values of backreaction and nonlinearity parameters are achieved.
In order to show that there is a good agreement between analytical
and numerical results, we present the numerical results in
tables I, II and III, too. However, we observe differences
in results in some cases which originates from the fact that in
order to solve the analytical solution, we use some
simplifications. One may argue that these disagreements could be
solved by considering the polynomial in the form of $F(z)=1-\alpha
z^2-\beta z^3 - \gamma z^4$, as the trial function. However, in
this case one faces with difficulties to achieve the solutions for
larger values of the BI and backreaction parameters. Indeed, in
this case, besides finding the parameter $\alpha$, we have to find
$\beta$ and $\gamma$ parameters. Actually, the analytical method
in holographic p-wave superconductors is very difficult. And for
this reason, most studies on the holographic p-wave
superconductors have been carried out numerically. In addition,
the investigation of one dimensional holographic superconductors
in the background of three dimensional BTZ black hole is a
difficult problem due to the logarithmic behavior of the gauge
field $\phi$. Overall, the shooting method's results follow the
same trend as the results of the Sturm-Liouville method, namely,
by increasing the strength of backreaction as well as the
nonlinearity parameters for each form of the trial
function $F(z)$. Indeed, increasing the value of backreaction and
nonlinear parameters, makes the condensation much harder to form.
In addition for $b=0$, the numerical values of \cite{mahyap} are
regained. Figures \ref{fig1} and \ref{fig3} show, respectively,
the behavior of condensation as a function of temperature for
different values of backreaction and nonlinear parameters. Based
on these figures, the condensation gap increases for larger values
of the backreaction and nonlinearity parameters, while the other
one is fixed. This implies that it is harder to form a holographic
$p$-wave superconductor in the presence of backreaction and BI
nonlinear parameters.
\section{Critical exponents}\label{sec4}
In this section, we are going to calculate the expectation value
of $\langle J_{x} \rangle$ in the vicinity of the critical
temperature $T_{c}$ for the one-dimensional holographic $p$-wave
superconductor developed in a BTZ black hole background, when the
gauge and vector fields backreact on the background geometry in
the presence of BI nonlinear electrodynamics. Again, we do our
calculations both analytically and numerically.
\subsection*{Analytical study}
To follow the analytical approach, we consider the behavior of
gauge field $A_{\mu}$ near the critical temperature. Since the
condensation in the vicinity of the critical temperature is
nonzero, we expect to have an extra term in the consequent
equation compared to the field equation (\ref{eqphiasym}) in the
previous section. Thus, Eq. (\ref{eqphiz}) turns to
\begin{equation}\label{eqphicrit}
\phi ''(z)+\frac{\phi '(z)}{z}+\frac{b z^3 \phi '^3(z)}{r_+^2}-\frac{2 \phi (z)
\rho _x{}^2(z)}{z^2 f(z)} \left(1-\frac{b z^4}{r_+^2} \phi '^2(z)\right)^{3/2}=0.
\end{equation}
Substituting Eqs. (\ref{eqfff}) and (\ref{eqrhoo}) in the above
expression, we get
\begin{equation}\label{eqphicritj}
\phi ''(z)+\frac{\phi '(z)}{z}+\frac{b z^3 \phi '^3(z)}{r_+^2}=
\frac{\langle J_x \rangle^2  }{r_+^4 } \phi (z) \Xi (z),
\end{equation}
where
\begin{equation}
\Xi (z)=\frac{ z^2 F^2(z) }{ g(z)} \left(1-\frac{3 b z^4 \phi
'^2(z)}{2 r_+^2}\right).
\end{equation}
In order to find the solution of Eq. (\ref{eqphicritj}), we note
that near the critical temperature, $T\simeq T_{c}$,  the value of
${\langle J_x \rangle^2 }/{r_+^4 }$ is small, so we may write the
solution in the form
\begin{equation}\label{eqphi2}
\frac{\phi (z)}{r_+}=\lambda  \log (z)-\frac{1}{4} b \lambda^3
\left(z^2-1\right)+\frac{\langle J_x \rangle^2 }{r_+^4}\eta (z).
\end{equation}
At the horizon $\phi(z=1)=0$ we have $\eta(1)=0$. Substituting Eq.
(\ref{eqphi2}) in Eq. (\ref{eqphicritj}) we arrive at
\begin{equation}\label{eqeta}
\eta''(z)+\frac{\eta '(z)}{z}+3 b z \lambda ^2 \eta
'(z)=\frac{\lambda  z^2 F^2(z)}{g(z)} \left[\log (z)-\frac{1}{4} b
\lambda ^2 \left(z^2-1\right)-\frac{3}{2} b \lambda^{2} z^2 \log
(z)\right].
\end{equation}
Multiplying both sides of Eq. (\ref{eqeta}) by factor $z
e^{\frac{3}{2} b \lambda ^2 z^2}$, we get
\begin{equation}
\int_0^1 d(z e^{\frac{3}{2} b \lambda ^2 z^2} \eta
'(z))=e^{\frac{3 b \lambda ^2}{2}}\eta '(1)=\lambda
\mathcal{A},\label{etap}
\end{equation}%
where
\begin{equation}
\mathcal{A}=\int_0^1\frac{z^3}{g(z)} F^2(z) e^{\frac{3}{2} b
\lambda ^2 z^2} \left[\log (z)-\frac{1}{4} b \lambda ^2
\left(z^2-1\right)-\frac{3}{2} b \lambda^{2} z^2 \log (z)\right]
\, dz.
\end{equation}
Now, we use the coordinate transformation $%
z\rightarrow Z+1$ in Eq. (\ref{eqphiiians}). Considering the fact
that the first term on the rhs of Eq. (\ref{eqphi2}) is the
solution of $\phi(z)$ at the critical point, and the second term
is a correction term, we have
\begin{equation}
\frac{\rho }{r_{+}}+\frac{\mu }{r_{+}}\log (1+Z)=\frac{\mu }{r_{+c}} \log (1+Z)-\frac{1}{4} b
 \left(\frac{\mu }{r_{+c}}\right) ^3 \left[(1+Z)^2-1\right]+%
\frac{\langle J_{x}\rangle ^{2}}{r_{+c}^{4}} \eta (1+Z),  \label{eqq0}
\end{equation}%
Then, by expanding the resulting equation around $Z=0$ we find
\begin{equation}
\frac{\rho }{r_{+}}+\frac{\mu }{r_{+}}\left(Z-\frac{Z^{2}}{2}+...\right)=\frac{\mu }{%
r_{+c}}\left(Z-\frac{Z^{2}}{2}+...\right)-\frac{1}{4} b  \left(\frac{\mu }{r_{+c}}\right) ^3 \left[(1+Z)^2-1\right]+\frac{\langle
J_{x}\rangle ^{2}}{r_{+c}^{4}}\left(\eta (1)+Z\eta
^{\prime }(1)+...\right). \label{eqq}
\end{equation}%
Comparing the coefficients $Z$ on both sides of Eq. (\ref{eqq})
and using Eq. (\ref{etap}) we find
\begin{equation}
\frac{\mu }{r_{+}}=\frac{\mu }{r_{+c}} \left( 1+\frac{\langle J_{x}\rangle
^{2}}{r_{+c}^{4}} \mathcal{A} e^{-\frac{3}{2} b \lambda ^{2}} \right) , \label{samez}
\end{equation}%
Near the critical point we have $T\sim T_{c}$, and thus using
relation (\ref{eqtemp}), we can find $r_{+}$ as
\begin{equation}
r_{+}=\frac{4\pi T}{\left( 2-\kappa ^{2}\lambda ^{2}+\frac{1}{4} b
\kappa ^2 \lambda ^4\right)}. \label{eqr+}
\end{equation}%
Inserting Eqs. (\ref{eqtemp}) and (\ref{eqr+}) in Eq. (\ref{samez})
and taking the absolute values of the resulting equation, we
arrive at
\begin{equation}
\langle J_{x}\rangle =\gamma T_{c}^{2}\sqrt{1-\frac{T}{T_{c}}}, \label{eqJx}
\end{equation}%
where
\begin{equation}\label{gamman}
\gamma = \frac{1}{\sqrt{\left\vert \mathcal{A}\right\vert }}\left(
\frac{4\pi }{2-\kappa ^{2}\lambda ^{2}+\frac{1}{4} b \kappa ^2 \lambda ^4}\right) ^{2} e^{\frac{3}{4} b \lambda ^2}.
\end{equation}
According to this equation, the critical exponent $\beta=1/2$ is
in good agreement with the mean field theory. We face with the
second order phase transition for all values of the backreaction
and nonlinear parameters because the value of $\beta$ is
independent of the effect of backreaction and nonlinearity. In
addition, the equation (\ref{gamman}) for $b\rightarrow 0$ turns
to equivalent equation in \cite{mahyap}.
\subsection*{Numerical approach}
\begin{figure*}[t]
\centering
\subfigure[~$  b=0$]{\includegraphics[width=0.3\textwidth]{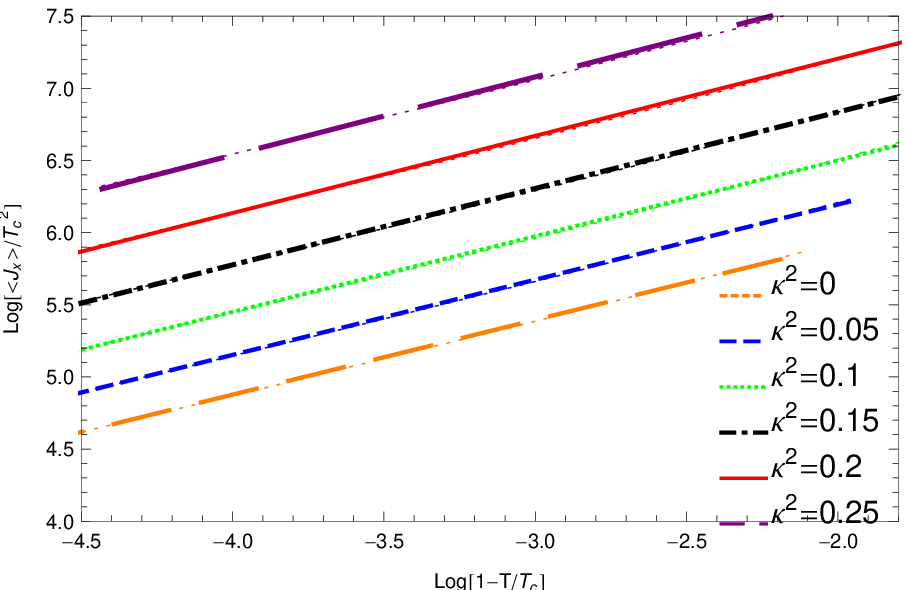}} \qquad %
\subfigure[~$  b=0.04$]{\includegraphics[width=0.3\textwidth]{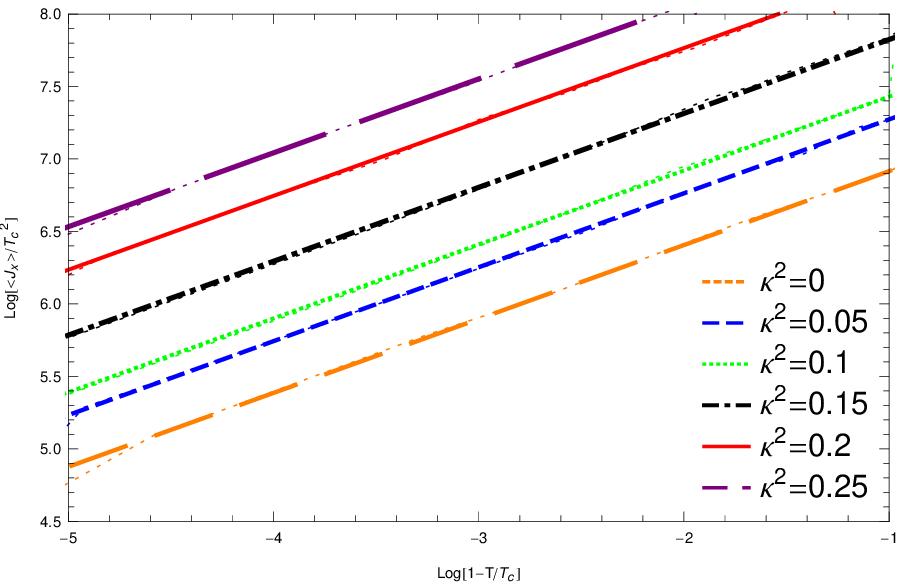}} \qquad %
\subfigure[~$  b=0.08$]{\includegraphics[width=0.3\textwidth]{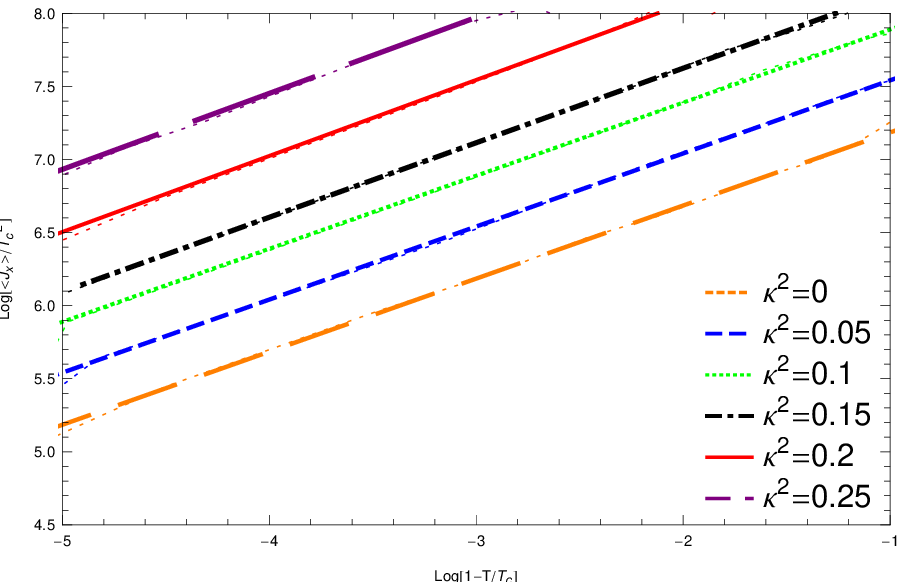}}
\caption{The behavior of $\log \langle J_x \rangle/T_c^2$ as a function of $\log(1-T/T_{c})$
with slope of $1/2$ for different values of backreaction parameters.}
\label{fig4}
\end{figure*}

\begin{figure*}[t]
\centering
\subfigure[~$  \kappa^{2}=0$]{\includegraphics[width=0.3\textwidth]{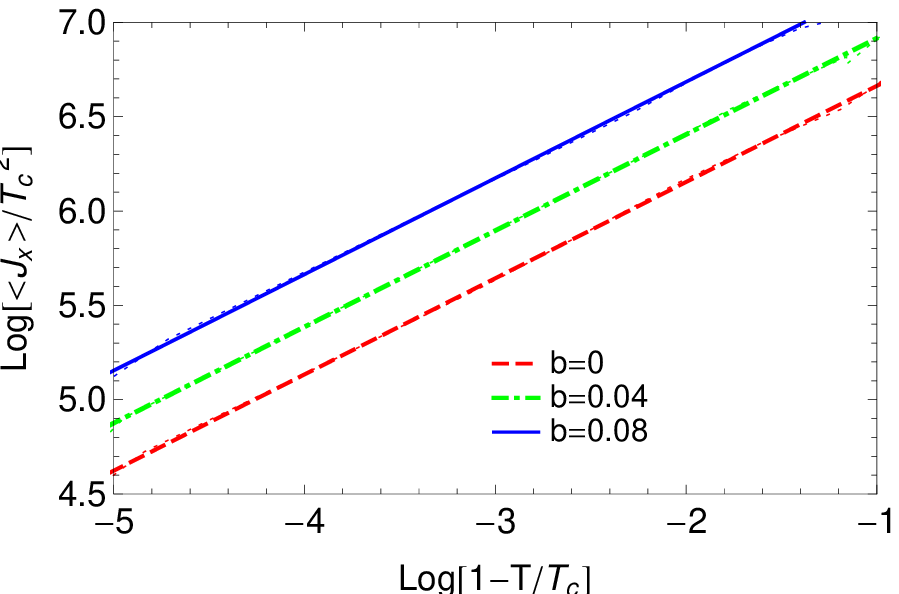}} \qquad %
\subfigure[~$  \kappa^{2}=0.10$]{\includegraphics[width=0.3\textwidth]{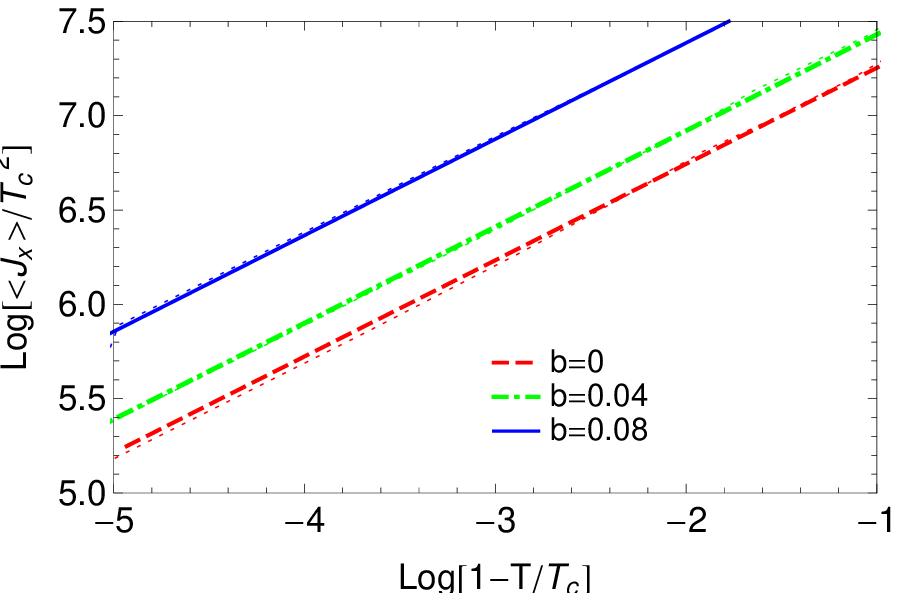}} \qquad %
\subfigure[~$  \kappa^{2}=0.20$]{\includegraphics[width=0.3\textwidth]{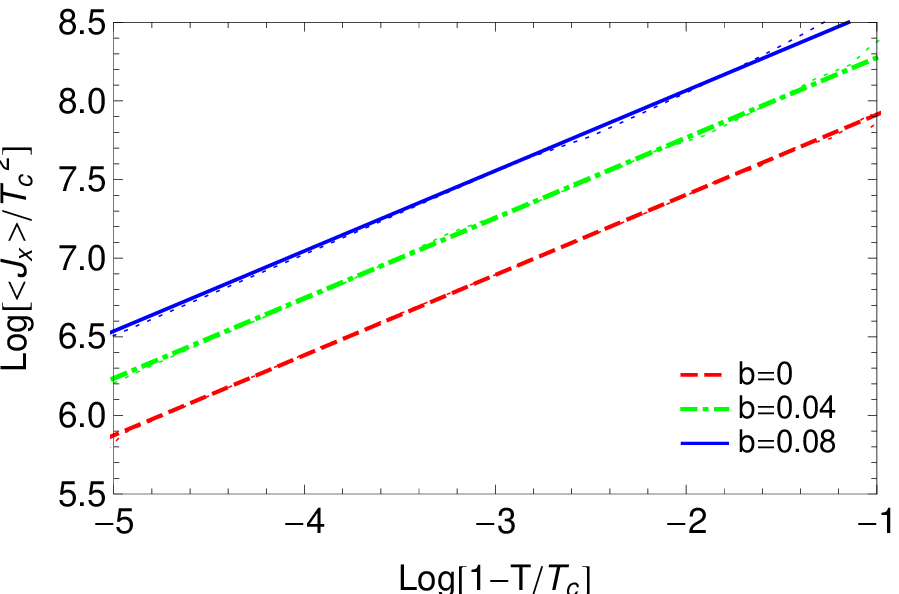}}
\caption{The behavior of $\log \langle J_x \rangle/T_c^2$ as a function of $\log(1-T/T_{c})$ with
slope of $1/2$ for different values of nonlinearity parameters.}
\label{fig5}
\end{figure*}

Using the results of analytical solution for the condensation in
the vicinity of the critical temperature (i.e. equation
(\ref{eqJx})) we have
\begin{equation}
\log \left(\frac{\langle J_x \rangle}{T_c^2}\right)=\log (\gamma )+\frac{1}{2} \log \left(1-\frac{T}{T_c}\right).
\end{equation}
Figures \ref{fig4} and \ref{fig5} give information about the
behavior of $\log \left(\frac{\langle J_x \rangle}{T_c^2}\right)$
as a function of $\log \left(1-\frac{T}{T_c}\right)$ in the
presence of backreaction and BI nonlinear parameters. The slope of
curves is $1/2$ which is in agreement with analytical approach. In
addition, both methods follow the mean field theory and the second
order phase transition is occurred.
\section{Conductivity}\label{sec5}
\begin{figure*}[t]
\centering
\subfigure[~$  b=0$]{\includegraphics[width=0.3\textwidth]{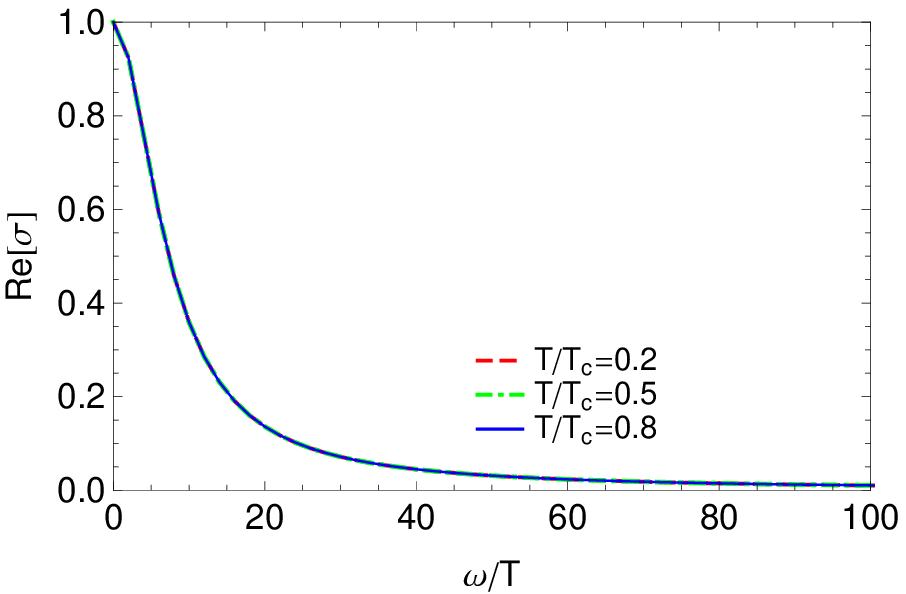}} \qquad %
\subfigure[~$  b=0.04$]{\includegraphics[width=0.3\textwidth]{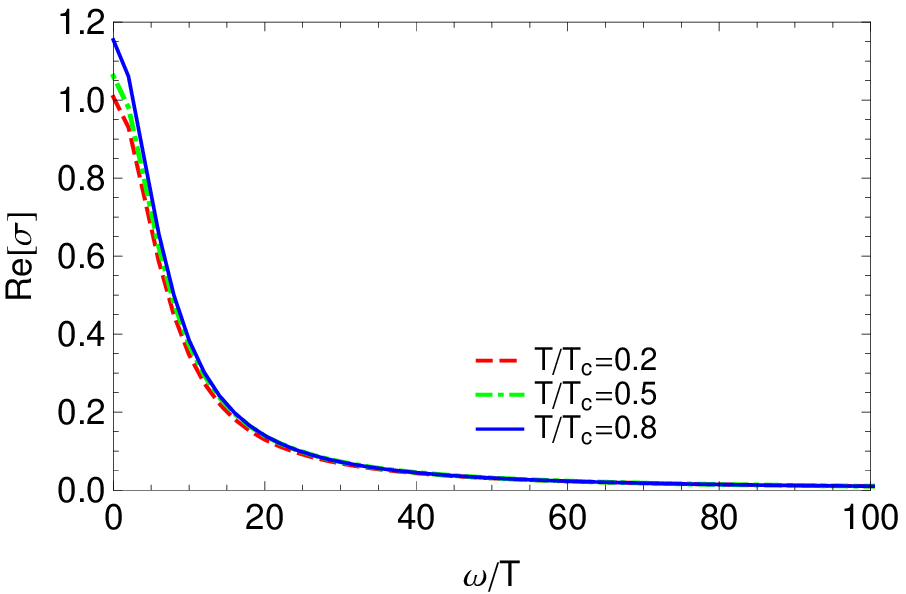}} \qquad %
\subfigure[~$  b=0.08$]{\includegraphics[width=0.3\textwidth]{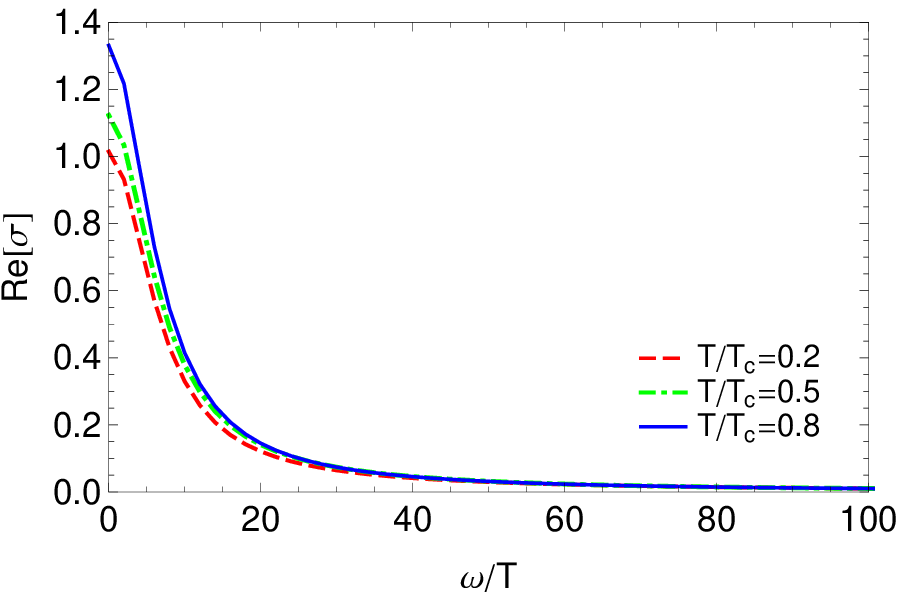}}
\caption{The behavior of real part of conductivity as a function of $\omega/T$ for different values of temperature in the case $\kappa^{2}= 0$.}
\label{fig6}
\end{figure*}

\begin{figure*}[t]
\centering
\subfigure[~$ b=0$]{\includegraphics[width=0.3\textwidth]{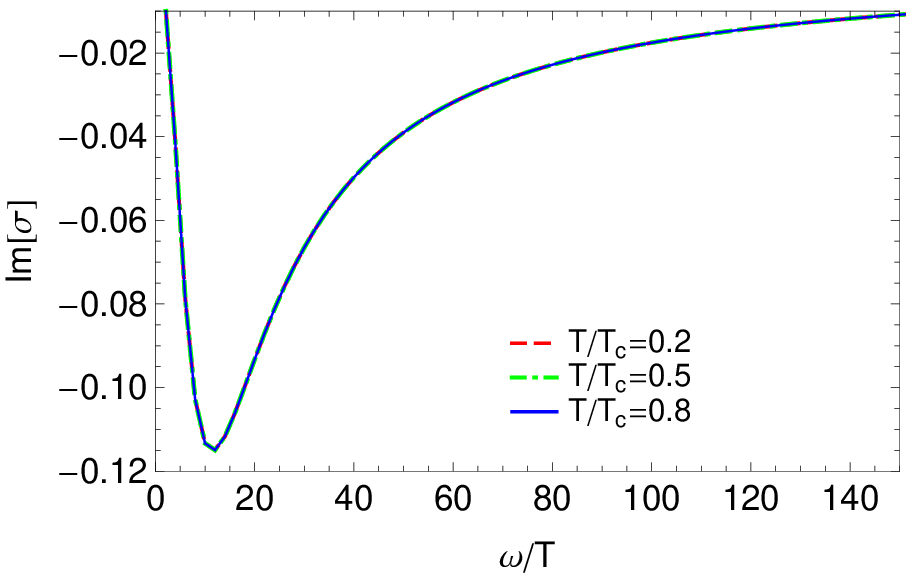}} \qquad %
\subfigure[~$ b=0.04$]{\includegraphics[width=0.3\textwidth]{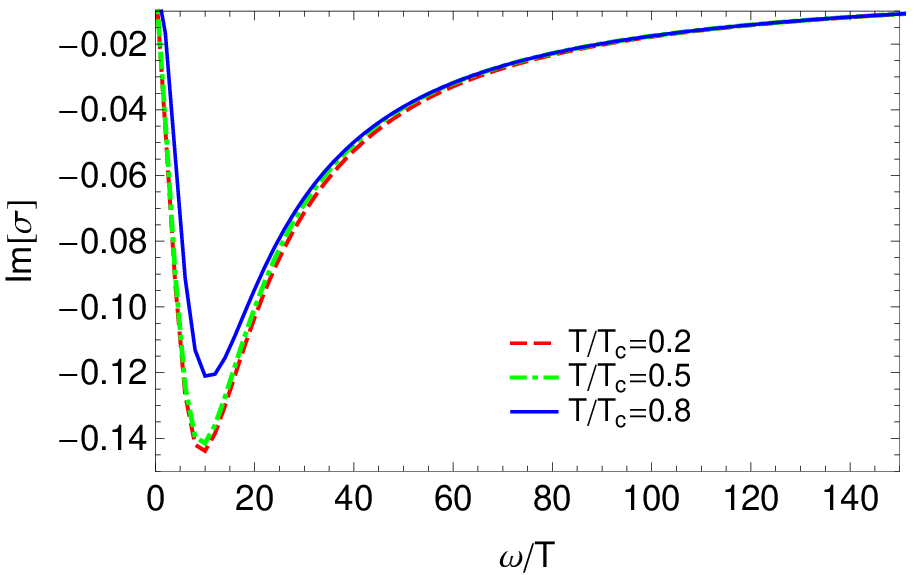}} \qquad %
\subfigure[~$ b=0.08$]{\includegraphics[width=0.3\textwidth]{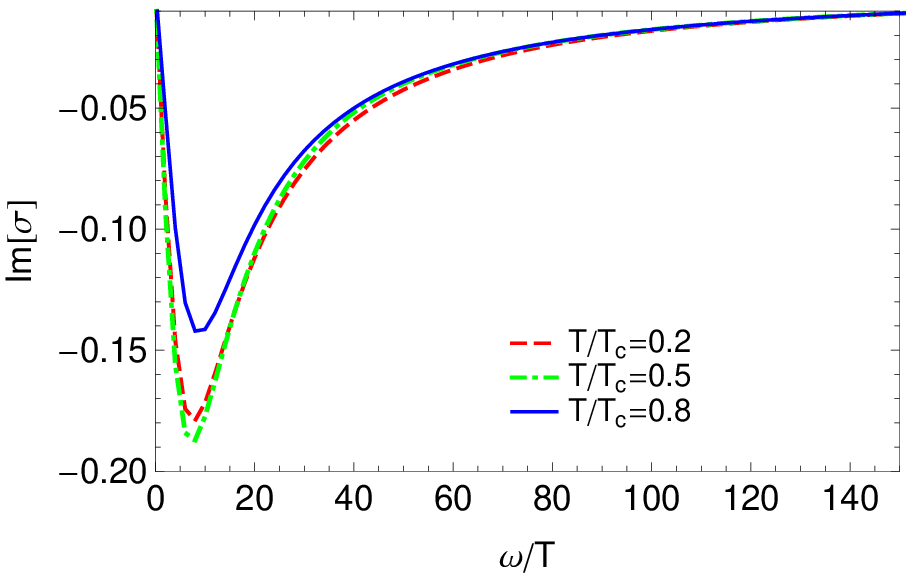}}
\caption{The behavior of imaginary part of conductivity as a function of $\omega/T$ for different values of temperature in the case $\kappa^{2}= 0$.}
\label{fig7}
\end{figure*}

\begin{figure*}[t]
\centering
\subfigure[~$  T/T_{c}=0.2$]{\includegraphics[width=0.3\textwidth]{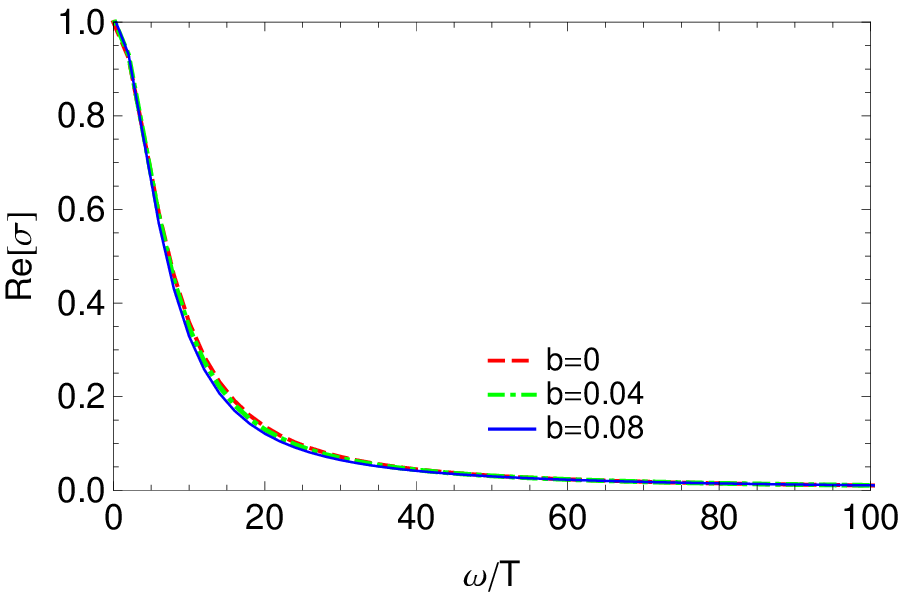}} \qquad %
\subfigure[~$  T/T_{c}=0.5$]{\includegraphics[width=0.3\textwidth]{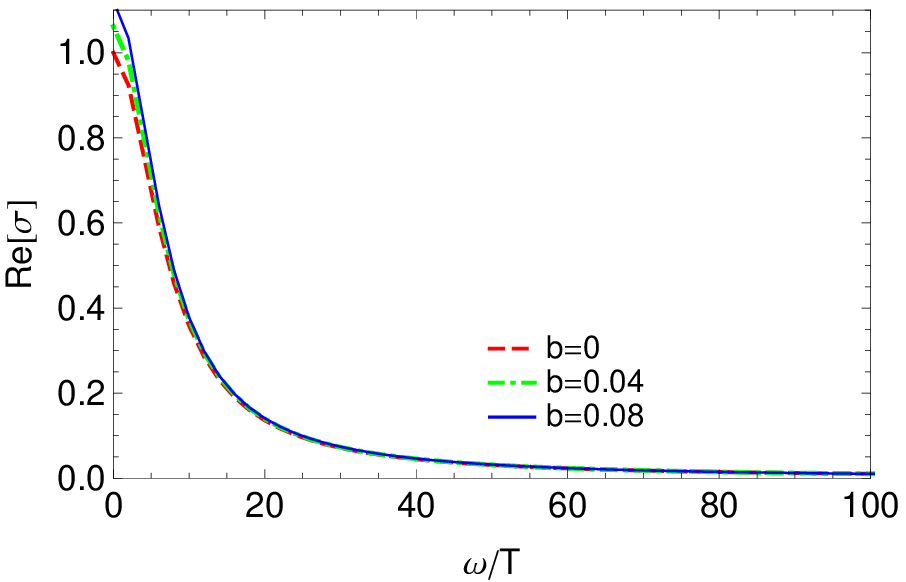}} \qquad %
\subfigure[~$  T/T_{c}=0.8$]{\includegraphics[width=0.3\textwidth]{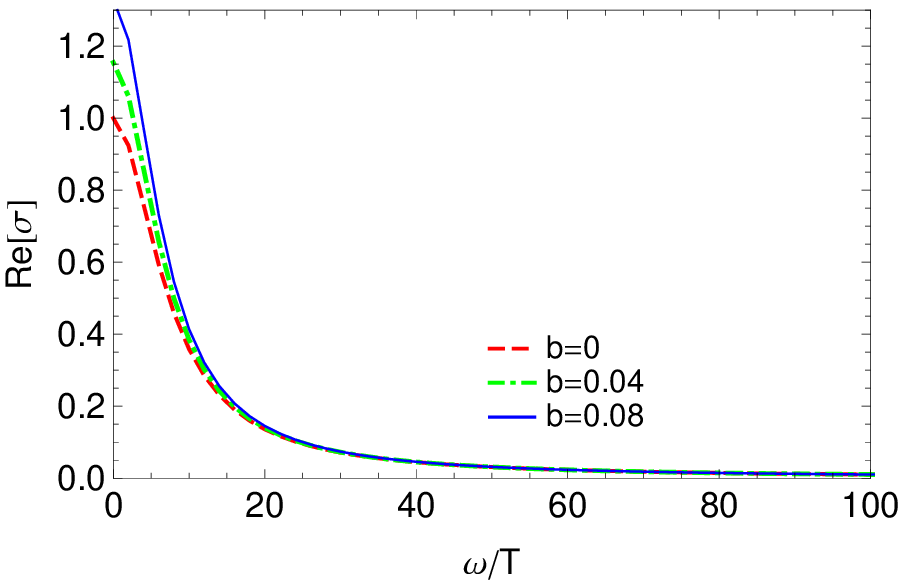}}
\caption{The behavior of real part of conductivity as a function of $\omega/T$ for different values of $b$ in the case $\kappa^{2}= 0$.}
\label{fig8}
\end{figure*}

\begin{figure*}[t]
\centering
\subfigure[~$  T/T_{c}=0.2$]{\includegraphics[width=0.3\textwidth]{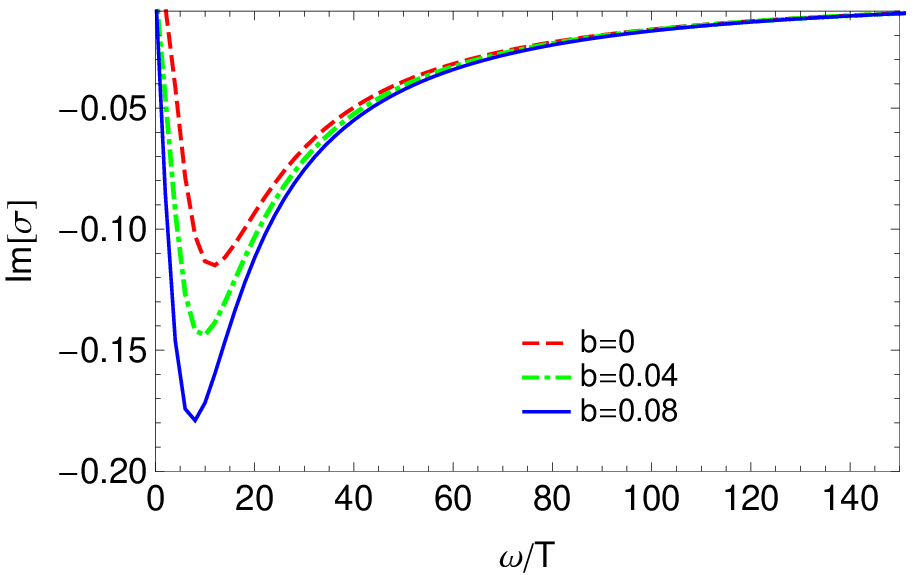}} \qquad %
\subfigure[~$  T/T_{c}=0.5$]{\includegraphics[width=0.3\textwidth]{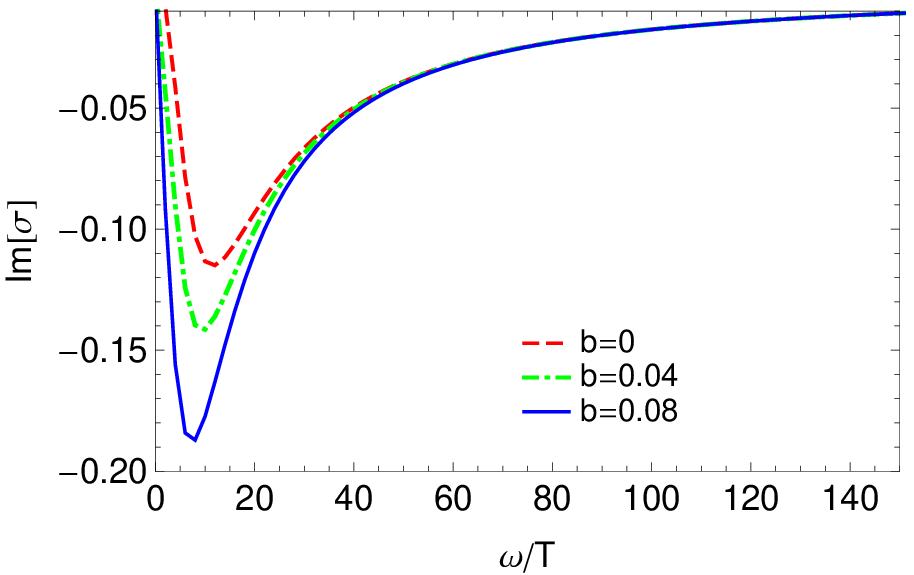}} \qquad %
\subfigure[~$  T/T_{c}=0.8$]{\includegraphics[width=0.3\textwidth]{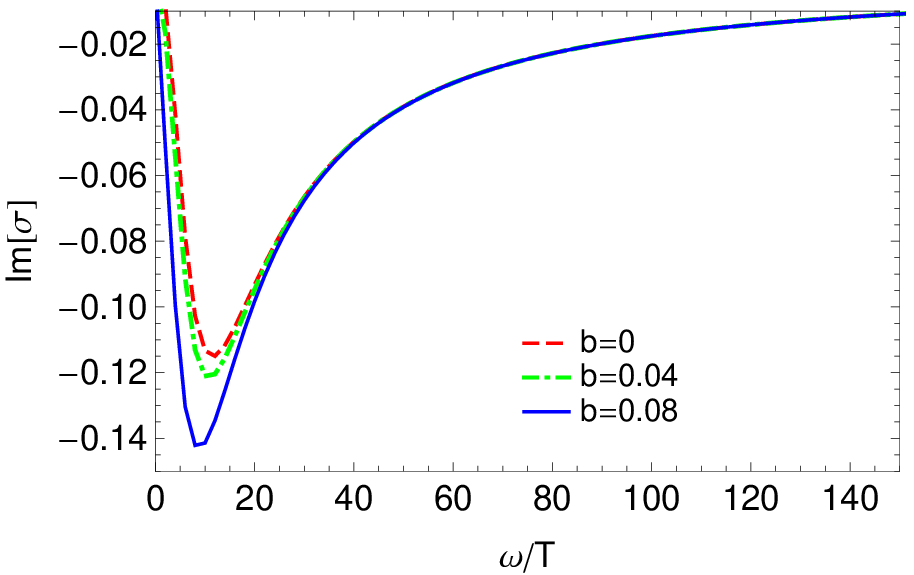}}
\caption{The behavior of imaginary part of conductivity as a function of $\omega/T$
for different values of $b$ in the case $\kappa^{2}= 0$.}
\label{fig9}
\end{figure*}

In this section, we obtain the electrical conductivity as a
function of frequency for the one-dimensional holographic $p$-wave
superconductors in the presence of backreaction and BI nonlinear
electrodynamics by applying appropriate electromagnetic
perturbations of $A_{x}$ and $g_{tx}$ on the black hole
background. Based on the AdS/CFT correspondence, these
perturbations in the bulk are dual to the boundary electric
current. If we consider $\sigma_{ij}$ and $J_{i}$ as the electric
conductivity and external electric field, according to the Ohm's
law we have
\begin{equation}\label{ohm}
\sigma_{ij}=\frac{J_{i}}{E_{j}}.
\end{equation}
In order to calculate the conductivity in $x$-direction we need to
add the following perturbational terms in the bulk gauge potential
and metric
\begin{equation}\label{perturb}
\delta A_{x}=A_{x}(r) e^{-i \omega t},\  \  \  \delta g_{tx}=g_{tx}(r) e^{-i \omega t}.
\end{equation}
Using Eq. (\ref{perturb}), the linearized form of $x$-component of
the electromagnetic equation (\ref{eqmax}) turns to
\begin{eqnarray}
A_{x}''(r)&+&A_{x}'(r)\left[\frac{f'(r)}{f(r)}-\frac{\chi '(r)}{2}-\frac{1}{r}+\frac{2 b \rho_{x} ^2 (r) e^{\chi (r)} \phi (r) \phi '(r) }{r^2 f(r)}\sqrt{1-b e^{\chi (r)} \phi '^2(r)}-\frac{b e^{\chi (r)} \phi '^2(r)}{r}\right]\notag \\
&+&A_{x}(r)\frac{\omega ^2 e^{\chi (r)}}{f^2(r)}+\frac{e^{\chi
(r)} \phi '(r)}{ f(r)} \left[g_{tx}'(r)-\frac{2
g_{tx}(r)}{r}\right]+\frac{2  \rho_{x} ^2(r) e^{\chi (r)} \phi (r)
}{r^2 f^2(r)}g_{tx}(r) \sqrt{1-b e^{\chi (r)} \phi '^2(r)}=0.
\label{eqcong}
\end{eqnarray}%
Also, using  $(tt),(xx),(tx),(xt),(xr)$-components of the Einstein
equations and after some simplification, we arrive at
\begin{equation}
\frac{1}{2 \kappa ^2}\left[g_{tx}'(r)-\frac{2 g_{tx}(r)}{r}\right]+\frac{A_{x}(r) \phi '(r)}{\sqrt{1-b e^{\chi (r)} \phi '^2(r)}}=0,\label{eq01}
\end{equation}
\begin{equation}
\frac{2 }{r^2}A_{x}(r) \rho_{x} ^2(r)+\frac{2 \rho_{x} ^2(r)
e^{\chi (r)} \phi (r)}{r^2 f(r)}g_{tx}(r) =0.\label{eq02}
\end{equation}
Substituting Eqs. (\ref{eq01}) and (\ref{eq02}) in
Eq.(\ref{eqcong}), we obtain the linearized equation for the gauge
field $A_{x}$
\begin{eqnarray}
A_{x}''(r)&+&A_{x}'(r)\left[\frac{f'(r)}{f(r)}-\frac{\chi '(r)}{2}-\frac{1}{r}+\frac{2 b \rho_{x} ^2(r) e^{\chi (r)} \phi (r) \phi '(r) }{r ^2 f(r)}\sqrt{1-b e^{\chi (r)} \phi '^2(r)}-\frac{b e^{\chi (r)} \phi '^2(r)}{r}\right]\notag \\
&+&A_{x}(r)\left[\frac{\omega ^2 e^{\chi (r)}}{f^2(r)}-\frac{2 \rho_{x} ^2(r) }{r^2 f(r)}\sqrt{1-b e^{\chi (r)} \phi '^2(r)}-\frac{2 \kappa ^2 e^{\chi (r)} \phi '^2(r)}{f(r) \sqrt{1-b e^{\chi (r)} \phi '^2(r)}}\right]=0. \label{eqconf}
\end{eqnarray}%
Let us note that, of course, investigating the effects of
nonlinearity as well as the backreaction parameters on the
conductivity is a worthy task. However in order to compute the
conductivity in the holographic approach in the presence of
nonlinear and backreaction parameters, we need to turn on the one
component of the gauge field as well as the metric. This extra
component makes the calculations too difficult and we face with
difficulty in numerical solutions for the cases with nonzero
values of backreaction. Therefore, for simplicity, in what
follows, we consider only the probe limit by setting
$g_{tx}(r)=0=\chi(r)$ in the presence of BI nonlinear
electrodynamics. Thus, in the absence of backreaction, Eq.
(\ref{eqcong}) can be written as
\begin{equation}
A_{x}''(r)+A_{x}'(r) \left[\frac{f'(r)}{f(r)}-\frac{1}{r}+\frac{2 b \rho_{x} ^2(r) \phi (r)
\phi '(r)}{r^2 f(r)} \sqrt{1-b \phi '^2(r)}-\frac{b \phi '^2(r)}{r}\right]+\frac{\omega ^2 A_{x}(r)}{f^2(r)}=0, \label{eqconfff}
\end{equation}%
where it admits the asymptotic solution in the following form
\begin{equation}
A_{x}=A_{x}^{(0)}+A_{x}^{(1)} \log \left(\frac{1}{r}\right). \label{eqaasymp}
\end{equation}%
Based on the AdS/CFT dictionary, $A_{x}^{(0)}$ plays the role of
the source in the dual theory, while $A_{x}^{(1)}$ will give the
expectation value of the dual current. For the boundary current we
have
\begin{equation}
J=\frac{\text{$\delta $S}_{\text{bulk}}}{\text{$\delta $A}^{(0)}}=\frac{\text{$\delta $S}_{o.s}}
{\text{$\delta $A}^{(0)}}=\frac{\partial(\sqrt{-g}\mathcal{L}_{m})}{\partial A_x'}\vert r\rightarrow \infty,
\end{equation}
where
\begin{equation}
S_{o.s.}=\int_{r_+}^{\infty }\, dr \int \, {d}^{2}x \sqrt{-g}\mathcal{L}_{m}.
\end{equation}
Integrating by parts and using Eq. (\ref{eqconfff}), we get
\begin{equation}
S_{o.s.}=\int \, {d}^{2}x \frac{f(r) A_x(r) A_x'(r)}{2 r \left(1-b \phi '^2(r)\right)}.
\end{equation}
Using the asymptotic behavior of $\phi(r)$, $f(r)$ and $A_{x}(r)$
given by Eqs. (\ref{eqasymp}) and (\ref{eqaasymp}), we can
calculate $J_{x}$. So, the electrical conductivity based on the
equation (\ref{ohm}) is
\begin{equation} \label{eqconductivity}
\sigma(\omega)=\frac{J_{x}}{E_{x}}=-\frac{i A_x^{(1)}}{\omega  A_x^{(0)}}=-\frac{i z A_x'(z)}{\omega  \left[A_x(z)-z A_x'(z) \log (z)\right]},
\end{equation}%
where
\begin{equation}
E_{x}=-\partial_{t} \delta A_{x}.
\end{equation}
Following the analytical approach to calculate conductivity seems
difficult thus we apply the numerical method. In order to do that,
we consider ingoing wave boundary condition in the vicinity of the
horizon for $A_{x}(r)$ as follow
\begin{equation}
A_{x}(r)=f(r)^{\frac{-i \omega}{4 \pi T}} \left[1+a(1-r)+b(1-r)+\cdots\right].
\end{equation}
In the above equation, $T$ is the Hawking temperature which in the
probe limit $T=r_{+}/(2 \pi)$ because in this case $ f(r) = r^2 -1
$. Furthermore, $a$, $b$, $\cdots$ are obtained based on the
Taylor expansion of equation (\ref{eqconfff}) around horizon. Now,
due to Eq. (\ref{eqconductivity}) we can numerically explore the
behavior of the conductivity for the $(1+1)$-dimensional
holographic $p$-wave superconductor in the probe limit in presence
of BI nonlinear electrodynamics. Figures \ref{fig6} and \ref{fig7}
give information about the behavior of the real and imaginary
parts of conductivity as a function of $\omega/T$ for different
values of nonlinearity parameter $b$ in the case $\kappa^{2}=0$
for $T/T_{c}=0.2, 0.5, 0.8$. Delta function in the real part of
the conductivity isn't related to imaginary part near $\omega/T=0$
by the Kramers-Kronig relation because imaginary part tends to
zero instead of having a pole in this region. In addition,
$Re\sigma(\omega)$ tends to zero value at high frequency same as
\cite{Bu}. The imaginary and real parts of conductivity follow a
different trend in higher dimensions because we face with the
absence of gap and divergence behavior in real and imaginary
parts, respectively. Moreover, the effect of different values of
temperature is more apparent for larger values of nonlinearity
parameter. Figures \ref{fig8} and \ref{fig9} show the effect of
different values of nonlinearity parameter $b$ for fixed values of
$T/T_{c}$. Based on these figures, the difference of graphs
becomes more obvious by increasing the value of $T/T_{c}$. In
addition, for the fixed value of $T/T_{c}$ strengthen the effect
of nonlinearity makes Drude-like peak in real part of conductivity
to increase and causes deeper minimum values in imaginary part.
\section{summary and discussion}\label{sec6}
We have investigated one-dimensional holographic $p$-wave
superconductor model by applying $AdS3/CFT2$ when the gauge and
vector fields backreact on the background geometry in the presence
of BI nonlinear electrodynamics. For this purpose, we employ the
Sturm-Liouville eigenvalue problem for analytical investigations
and the shooting method for the numerical calculations. In both
methods, we find the relation between the critical temperature
$T_{c}$ and the chemical potential $\mu$ for different values of
the nonlinear and backreaction parameters. The results of
analytical and numerical methods are in good agreement with each
other. We found out that increasing the values of the nonlinearity
and backreaction parameters decrease the critical temperature and
thus makes the condensation harder to form. Furthermore, critical
exponent of this system were also obtained both analytically and
numerically. We face with a second order phase transition with
$\beta=1/2$ which follows the mean field theory. This value is
independent of backreaction and nonlinear effects.

In addition, we analyzed the conductivity of this system for the
case of probe limit and investigated the properties of real and
imaginary parts of conductivity for different values of the
nonlinear parameter $b$. The behavior of both real and imaginary
parts of conductivity are so different from higher dimensions and
they don't connect to each other based on the Kramers-Kronig
relation. We don't observe divergency near $\omega/T=0$ in the
imaginary part of the conductivity where a delta function in the
real part appears. By increasing the effect of nonlinearity, we
obtain larger values for Drude-like peak in real part of
conductivity and deeper minimum values of imaginary part. It's
difficult to see the effect of different temperatures for small
values of nonlinearity parameter. However, the effect of
temperature becomes apparent by increasing the $b$.
\begin{acknowledgments}
We are grateful to the referee for constructive comments which
helped us improve our paper significantly. We also thank Shiraz
University Research Council. The work of AS has been supported
financially by Research Institute for Astronomy and Astrophysics
of Maragha (RIAAM), Iran.
\end{acknowledgments}

\end{document}